# A methodical approach to evaluate the potential of Quantum Computing for Manufacturing Simulations

S. Schröder[a]*, J. Felipe[a], S. Danz[b,c], P. Kienast[a], A. Ciani[b], P. Ganser[a], T. Bergs[a,d]

[a]Fraunhofer Institute for Production Technology IPT, Steinbachstraße 17, Aachen 52074, Germany
[b]Institute for Quantum Computing Analytics (PGI-12), Forschungszentrum Jülich, 52425 Jülich, Germany
[c]Theoretical Physics, Saarland University, Saarbrücken 66123, Germany
[d]Manufacturing Technology Institute - MTI of RWTH Aachen University, Campus-Boulevard 30, Aachen 52074, Germany

* Corresponding author: stefan.schroeder@ipt.fraunhofer.de

In this paper, a methodical approach to evaluate the potential of quantum computing for manufacturing simulation, using the example of multi-axis milling of thin-walled aerospace components, is discussed. A developed approach for identifying bottlenecks in manufacturing simulations, for which the application of quantum computing potentially provides a speed-up or increase in accuracy, is presented. Moreover, indicators of quantum computing suitability and feasibility are defined with the main objective of identifying whether a manufacturing simulation bottleneck is suitable for quantum computing applications. First results of testing a hybrid routine as an application approach for the milling dynamics simulation on quantum machines are presented.

## 1. Introduction and Motivation

It is currently difficult for industrial companies to assess the topic of quantum computing (QC) and its potential for their production. The first steps in estimating which problems are likely to be solved more efficiently by a quantum computer and which have QC-potential can be found in literature. [1-4] Classifying use cases in the field of QC along three dimensions (application domain, problem category, and problem class) makes it easier to categorize them in the overall landscape and this enables users to identify similar use cases. Based on [5], Fig. 1 was created with a focus on the application domain "production". Industry-relevant problem categories, in which QC is likely to play a role, are simulations, optimization problems, machine learning applications, and cryptography [5]. Focus of this paper will be on the problem category simulation, explicitly manufacturing simulation. In the following, using a milling dynamics simulation, a methodical approach to evaluate the potential of QC for manufacturing simulations is presented. Section 2 presents algorithm-based (2.1) and hardware-related (2.2) indicators. Moreover, a methodical approach for the identification of QC-potentials in manufacturing simulations is presented in Section 3. In Section 4 the method is used to identify the QC-potential for a Finite Element (FE) modal analysis of a milling dynamics simulation. Moreover, in subsection 4.1 a hybrid routine as a preparation for a QC based modal analysis is presented, which has the potential to run on noisy intermediate-scale quantum (NISQ) devices. This hybrid routine will be referred to as the NISQ eigenvalue estimator. It was benchmarked at the IBM Quantum System One[1] at Ehningen Germany and the results are shown in subsection 4.2. The paper concludes with a summary of the results and an outlook in section 5.

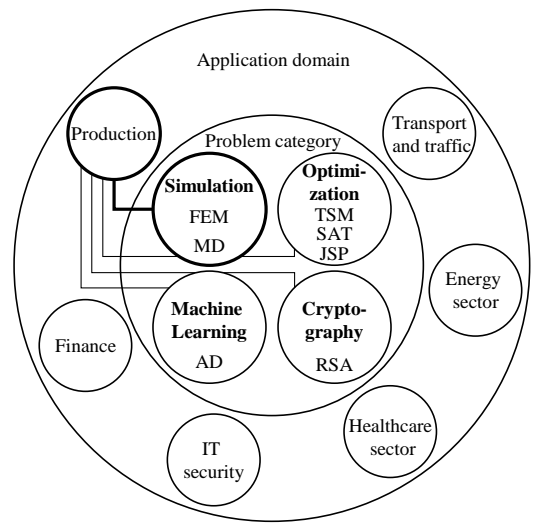

Fig. 1. Classification of production use cases along the three dimensions of the application domain, problem category, and problem class, based on [5]. The following problem classes are associated with the abbreviations and are examples of the respective category: FEM: Finite Element Method; MD: Molecular dynamics; AD: Anomaly detection; TSM: Traveling salesman problem; SAT: Satisfiability problem; JSP: Job scheduling problem; RSA: Rivest–Shamir–Adleman encryption

---

[1] IBM Quantum System One at Ehningen Germany: Number of qubits: 27; Coherence time ≈150 μs; Single qubit gate error ≈ 0.025 %; Two qubit gate error ≈ 0.7 %; Quantum Volume: QV 64

## 2. Definition of relevant evaluation indicators

To create an approach for directly identifying and determining the QC potential of manufacturing simulations, it was necessary to define indicators capable of executing this analysis and providing relevant and usable information. Such indicators will be referred to as **Quantum Potential Indicators (QPIs)**, instead of the standard Key Performance Indicator (KPI) usage. KPIs are measurable indications of performance over time for a specific goal. Although the objective of a set of metrics is to provide a quantified evaluation of quantum potential, these parameters are not always related to performance over time. A simulation potential for a QC-application depends primarily on its compatibility with quantum algorithms, i.e. if the problem can be tackled by quantum methods. The QPIs will be divided into software and hardware-related indicators. Simulation parameters and classical computing alternatives shall also be encompassed for the total completion of the method.

### 2.1. Software-related QPIs

Software-related QPIs are related to quantum algorithms, the problems they solve, and the conditions they require to be run. Since the focus of this paper is the milling dynamics simulation of thin-walled compressor blades and its linear systems of equations, the attention will be directed to investigating the Harrow, Hassidim, and Lloyd (HHL) [6] algorithm for the resolution of such linear systems, defined as: $A\vec{x} = \vec{b}$. Moreover, considering that the Quantum Phase Estimation (QPE) [7-10] is used in the HHL algorithm, both methods will have their requirements listed for further use as QPIs.

**HHL Algorithm**: The HHL algorithm, is a quantum algorithm to estimate features of the solution of a set of linear equations. It has five components: state preparation, QPE, ancilla bit rotation, inverse quantum phase estimation (IQPE), and measurement. The matrix $A$ must be hermitian, which guarantees that its eigenvalues are real numbers with eigenvectors that form an orthonormal basis. Moreover, this algorithm requires to prepare the vector $\vec{b}$ in a quantum state and implement $\sin^{-1}(C/\lambda_z)$ in a reasonable time. Here, $C$ is a normalization factor and $\lambda_z$ an eigenvalue of $A$. Another limitation of the HHL algorithm is its probabilistic nature. Not every measured result is correct and wrong results must be discarded. This can lead to a non-negligible overhead in sample size.

**QPE**: This routine allows to estimate the phase $\varphi$ applied by a unitary operator $U$ to a quantum state $|\Psi\rangle$: $U|\Psi\rangle = e^{2\pi i\varphi}|\Psi\rangle$. This can be used as eigenvalue estimator if one has access to a quantum gate of the form $U = e^{2\pi iA}$. Hence, the first requirement is that $U$ is unitary ($U^\dagger U = 1$) and can be implemented on a quantum computer. Also, not only this algorithm has specific requirements, but it also faces limitations. It requires a set of $m$ ancilla qubits to accurately estimate the phase up to $m$ bits. Moreover, a total of $2^m - 1$ executions of $U$ are necessary to prepare the phase in the ancilla qubits. Therefore, this algorithm scales exponentially with $m$.

### 2.2. Hardware-related QPIs

The quantum algorithms are executed on quantum hardware and their performance indicators are therefore treated as QPIs. As pointed out by Wack et al. [11], there are three main attributes in QC performance: scale, quality, and speed. These parameters provide the potential of a quantum computer to run real applications. For each one of these attributes, a metric will be used as a QPI, based on Wack's work. Wack et al. also created an illustrative benchmarking pyramid for speed and quality, in which they depicted a multilevel analysis structure with various indicators [11].

- **Scale:** it is based on the number of qubits and determines how much information can be encoded in a quantum system, therefore indicating the size of tractable problems [11-13]. To increase the number of qubits whilst maintaining acceptable coherence rates is a great challenge.
- **Quality:** it is measured by quantum volume (QV) which indicates the reliability level of a quantum circuit implementation [11, 12]. This parameter provides a holistic view of the system, as it takes into account coherence, gate fidelity, and measurement fidelity. Moreover, the parameter is influenced by compilers and connectivity.
- **Speed:** it is measured in circuit layer operations per second (CLOPS) and indicates the number of QV circuit layers that can be executed by a quantum processing unit per unity of time [11]. The CLOPS metric is a sort of equivalent to the clock rate in classical computers. It encompasses all relevant times during execution.

Later on, IBM developed two new metrics to update how quantum speed and quality are measured, namely the Error per Layered Gate (EPLG) and the CLOPS$_h$ [14]. These parameters come to satisfyingly encompass 100+ qubit processors' performance and support the utility-scale QC era development. Additionally, the necessity of such metrics arose from the limitation of simulating QV experiments for large enough systems [14].

- **EPLG:** This metric evaluates quantum processor performance by measuring gate-level errors. It begins with layer fidelity calculations, based on randomized benchmarking techniques [15, 16], to assess the fidelity of individual layers in a linked set of qubits. Each layer is carefully constructed to ensure that each qubit experiences a maximum of one two-qubit gate. By multiplying the fidelity values of these layers, the final layer fidelity is obtained. The EPLG is then calculated as $EPLG = 1 - (layerfidelity)^{(n_{2Q})^{-1}}$, where $n_{2Q}$ is the number of two-qubit gates (typically $n_{2Q} = N - 1$ for a linear chain of qubits, with $N$ is the number of qubits), providing insights into the overall device performance and gate-level data. [14, 17]

- **CLOPS$_h$:** The traditional interpretation of QV layers in Qiskit is being reconsidered due to discrepancies between theoretical and practical hardware implementation. IBM now employs CLOPS$_h$, a metric that accounts for actual hardware performance, in place of measuring virtual circuit layer operation per second (CLOPS$_v$). CLOPS$_h$ defines a layer as containing only two-qubit gates that can be executed in parallel on the system architecture, leading to a more accurate representation of hardware capabilities. This updated metric enables direct comparisons between quantum devices and improvements in error mitigation techniques. [14]

IBM also provides information about their devices regarding the parameters previously covered, displayed in Table 1. It is possible to see from this table that a larger computer doesn't necessarily mean a faster one, indicating that single-dimension benchmarks are not enough for accurately describing a system's performance. The threefold benchmark approach, in this context, satisfyingly comprises various necessary aspects for good performance user applications [11].

Table 1. Overview of current IBM devices (QV: Quantum Volume, CLOPS$_v$: virtual circuit layer operations per second, CLOPS$_h$: hardware-aware circuit layer operations per second, EPLG: Error per layered gate for a 100-qubit chain)

| Device | Qubits | QV | CLOPS$_v$ | CLOPS$_h$ | EPLG |
|---|---|---|---|---|---|
| ibm_nairobi | 7 | 32 | 2.6K | - | - |
| ibm_cairo | 27 | 64 | 2.4K | - | - |
| ibm_hanoi | 27 | 64 | 2.3K | - | - |
| ibm_algiers | 27 | 128 | 2.2K | - | - |
| ibm_sherbrook | 127 | 32 | - | 5K | 1.7% |
| ibm_brisbane | 127 | - | - | 5K | 1.9% |
| ibm_kyiv | 127 | - | - | 5K | 2.1% |
| ibm_quebec | 127 | - | - | 5K | 2.3% |
| ibm_kawasaki | 127 | - | - | 5K | 2.4% |
| ibm_osaka | 127 | - | - | 5K | 2.8% |
| ibm_cleveland | 127 | - | - | 5K | 2.9% |
| ibm_nazca | 127 | - | - | 5K | 3.2% |
| ibm_kyoto | 127 | - | - | 5K | 3.6% |
| ibm_cusco | 127 | - | - | 5K | 5.9% |
| ibm_torino | 133 | - | - | 3.8K | 0.8% |

## 3. Methodical approach

The QPIs presented can be used to evaluate the potential of QC for a specific simulation task. The first step is to determine whether a simulation step is a bottleneck, and then perform a three-step analysis (see Fig. 2):

- *Is the bottleneck QC-suitable? - If it is, is there a quantum algorithm that is a good fit for the given bottleneck?*
- *Is there a method capable of solving it without a big effort on classical hardware?*
- *Is this quantum improvement feasible (on current quantum devices)?*

To analyze QC-suitability, the two indicators *system size* and *number of steps* are used. Generally, the computation time on classical computer scales at least as $\mathcal{O}(N^3)$ (cf. p.474 in [18]) to solve $N$ coupled linear equations [6]. QC applications do have the potential of solving linear equations polynomial in $N$. Therefore, cases with bigger $N$ favor QC applications. However, this depends as well on the problem and the quantum algorithm. In parallel to that, the number of steps indicate how often calculations are executed in a simulation, depending on the chosen step size. Smaller step sizes result in higher accuracy but bigger computational overhead, increasing the overall number of steps [19] The larger this number, the greater the simulation's potential for QC-application, when considering dependent steps such as an iterative mesh refinement in a FEM simulation. Here, the analysis is done based on the degree of occurrence of such attributes. After QC-suitability is confirmed, it is possible to proceed to the following method's steps. The bottleneck will be evaluated to determine if a classical computing method provides a higher effective speed-up than a quantum computer's. As for the classical computing alternative, the method of parallelization is addressed [20, 21]. Applications involving several independent tasks or events happening concurrently at a high level of complexity are appropriate for parallel computing. Since in some cases, implementing a quantum method results in a loss of effective speed-up due to imperfect operations, a classical method with higher effective speed-up will be preferred. On the other hand, the implementation effort of a QC method can be estimated by evaluating the number of operations to be performed on the problem to satisfy the algorithm's implementation requirements. Algorithm platforms such as PlanQk can also be used to help getting an overview and a pool of existing algorithms and their implementation in Qiskit [22]. However, a one-to-one comparison between classical and quantum computation times is currently difficult due to variations in gate operation times between quantum computers. It is expected that these variations will improve drastically in the upcoming years as today technology is still in its early stages. After completing previous analyses, the bottleneck's QC-feasibility is addressed. Here, the QPIs defined in section 2.2 in the context of NISQ devices are used. The purpose of this evaluation is to define for which problem size and complexity the solution can be implemented and how useful the generated results are, given the limitations of these devices, such as the available number of quality qubits, small decoherence times, maximum circuit depth, and high error rates. Finally, once the general methical approach is defined, along with the indicators and their use, their application becomes possible. In this paper the application will be directed to manufacturing simulations and their steps, specifically those of the milling dynamics simulation of thin-walled turbomachinery components.

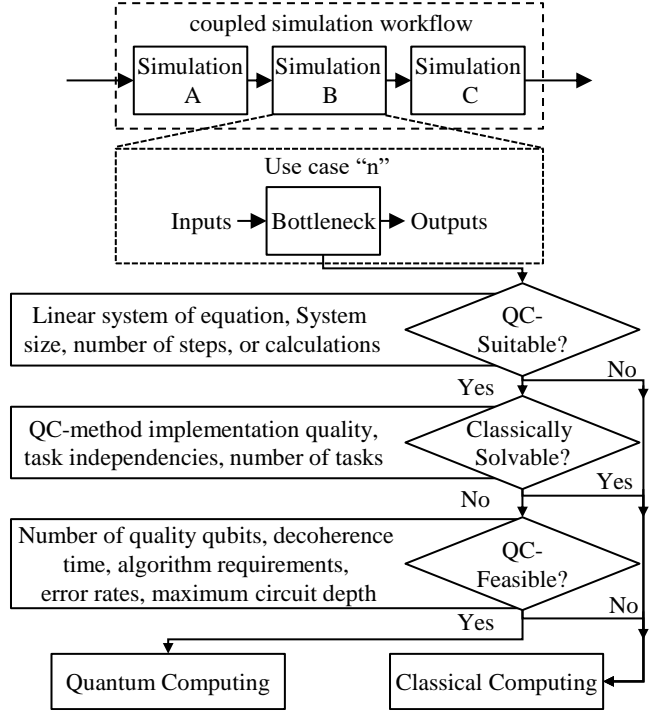

Fig. 2. Method schematic description (QC: Quantum Computing)

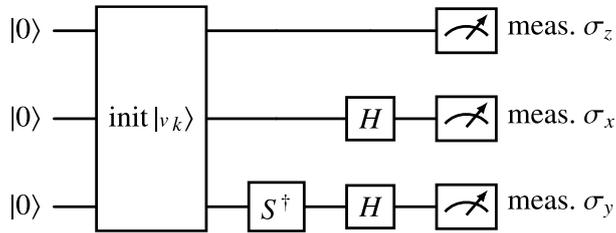

Fig. 3. Exemplary three-qubit circuit for step 3 of the full algorithm. The black box on the left initializes the quantum register by amplitude encoding $v_k$

## 4. FE-Modal analysis as first application approach

The optimization of the machining process of complex aerospace components, such as integral compressor rotors (blade integrated disks), is an active area of research due to their high cost. Process simulation is used to achieve optimization. Additionally, the advent of QC has brought a new tool with high optimization potential into focus. Schröder, Kienast et al. [23] applied the presented method to a milling dynamics simulation. The simulation is analyzed, and each component is evaluated separately concerning QC. The QPE is proposed to enhance the finite-element simulation-based component, such as the modal analysis for dynamics simulation. The paper presents the first investigations and results for the QC-based modal analysis of coupled oscillators. The results indicate that only steps 1 and 2 of the method were fulfilled. Critical voices could claim that a modal analysis could also be solved very efficiently in a classical way. However, it must be taken into account that in the use case of milling dynamics simulation, several million modal analyses must be carried out for a component so that classical methods reach their computational limits. To satisfy step 3, it is necessary to reduce the QPE approach to a minimum hybrid routine, which is introduced in the next section. However, this process results in the loss of the potential quantum advantage. [23] The following sections present the approach of using different geometries of coupled oscillators and benchmarking the results on a superconducting quantum computer from IBM[1].

*4.1. NISQ eigenvalue estimator*

State-of-the-art quantum computers are error-prone [24]. This limits the depth of quantum algorithms to small numbers of quantum operations before randomness dominates the system. Quantum algorithms with large numbers of quantum gates such as the HHL algorithm or the QPE are therefore not suitable for current devices. However, in the following, a minimal hybrid routine is described which is able to estimate the expectation value of a matrix $H$ on a NISQ device. This should provide insight into the feasibility of more complex algorithms. The measurement of quantum superposition leads to a collapse of the wave function into a random state. If one repeats the generation of this superposition followed by a measurement one can use this to estimate the expectation value $E(P)$ of the measured observable $P$. This makes them natural eigenvalue solver. For a quantum computer, those observables are, typically, the Pauli matrices $\sigma_i \in \{\sigma_x, \sigma_y, \sigma_z\}$ and their tensor products. An arbitrary matrix $H$ of size $N \times N$ can be decomposed into a superposition with those observables as basis

$$H = \sum_{i=1}^{N^2} g_i P_i, \qquad g_i = \frac{1}{N}\text{tr}(P_i H). \tag{1}$$

This allows us to estimate the expectation values of the single observables $P_i$ for a given eigenvector $v_k$ and construct the corresponding eigenvalue $\lambda_k$ of $H$ by summing over them with the corresponding weights $g_i$

$$\lambda_k = E_k(H) = \sum_{i=1}^{N^2} g_i E_k(P_i). \tag{2}$$

The quantum part of the algorithm is reduced to the amplitude encoding of $v_k$ in a quantum register followed by a measurement in the corresponding Pauli-base as illustrated in Fig. 3. The full routine for the computation of the eigenvalues $\lambda_k$ consist of four steps:

1. Compute the eigenvectors $v_k$ of $H$ on classical hardware.
2. Compute $g_i, \forall i \in \{1,2,...,N^2\}$ on classical hardware.
3. For all $P_i \in \{P_1, P_2, ... P_{N^2}\}$, encode the eigenvector $v_k$ in the amplitudes of a quantum register with $\lceil \log_2 N \rceil$ qubits and measure $P_i$ multiple times.
4. Sum over the the sampled expectation values $E_k(P_i)$ with the weights $g_i$ to have $\lambda_k$.

The most critical part of this routine is the amplitude encoding of the $N$ entries of $v_k$ as this requires to execute $\mathcal{O}(N)$ rotation gates on a quantum computer in the worst case. The more gates applied, the stronger the effect of the noise. This means that the accuracy of the eigenvalue estimations decrease with the size of the matrix $H$. The algorithm was tested by the authors on a superconducting quantum computer from IBM[1]. For this, a system of coupled oscillators for different geometries was considered and their resonance frequencies $\omega_k = \sqrt{\lambda_k}$ were computed; To be more precise, one-dimensional chains of coupled oscillators of various length and three different 3d-toy-models of compressor blades were investigated (see Fig. 4). The total number of oscillators in all those systems were kept small to minimize the effect of the noise. When computing the resonance frequencies on noisy quantum devices only an estimation $\tilde{\lambda}_k$ is achieved which lies between the exact eigenvalue $\lambda_k$ and the expectation value of the fully mixed or noise dominated state $\lambda_{\text{mixed}}$.

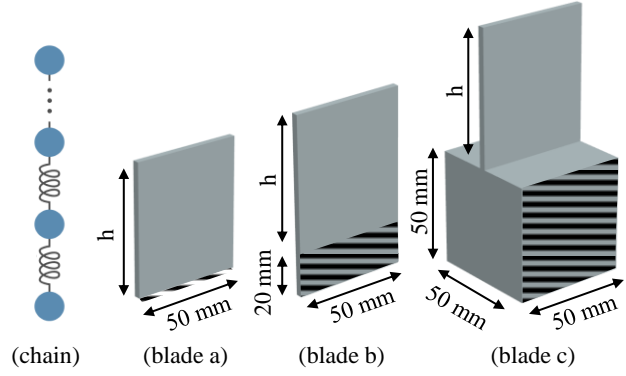

Fig. 4. Investigated simplified blade geometry: A chain of 2-64 oscillators (1-6 qubits), one blade geometry with 12 oscillators (blade a, 4 qubits) and two blade geometries with 24 oscillators allocated to every node and element of an arbitrary mesh (blade b and c, 5 qubits). The shaded areas represent fixed boundary conditions which reduced the total number of oscillators. The blade height $h$ is varied between $h = 10$ mm and $h = 60$ mm in 10 mm steps while keeping the oscillator number constant.

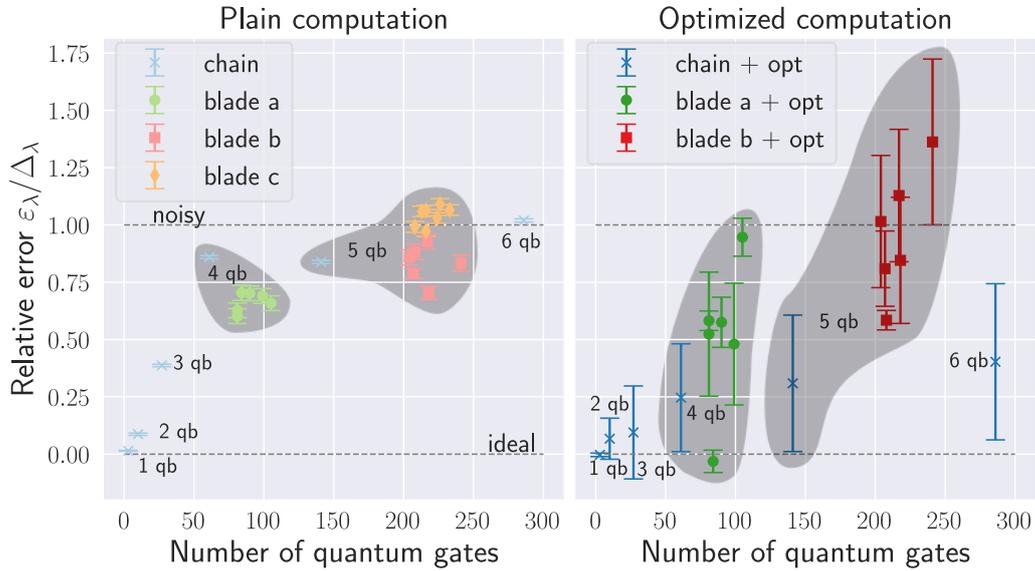

Fig. 5. Investigated relative eigenvalue error $\varepsilon_\lambda \Delta_\lambda^{-1}$ for different geometries (see Fig. 4) and oscillator numbers. In the left plot, the plain computation on a noisy quantum machine is shown. On the right state-of-the-art circuit optimization is used and error mitigation methods[2] are provided by IBM's Qiskit Primitives. The error bars in here represent statistical errors over 100 samples for chain and 1000 samples for blade geometries. Systems with more than 6 qubits (= 6 qb) were fully noise dominated in the tests.

The eigenvalue of the fully mixed state is equal to $g_1$ which corresponds to $P_1 = I_N$, where $I_N$ is the $N \times N$ identity matrix. All other contributions ($E(P_i) = 0, \forall i \neq 1$) vanish for the fully mixed state. In order to compare different geometries, focus will be on the error $\varepsilon_\lambda = \lambda_{\max} - \tilde{\lambda}_{\max}$ and a re-normalization of it with the width $\Delta_\lambda = \lambda_{\max} - \lambda_{\text{mixed}}$. Here the authors chose to concentrate on the maximum eigenvalue $\lambda_{\max}$ which allows to estimate the maximum errors as $\lambda_{\max} > \lambda_{\text{mixed}}$ for the described problems. The relative error $\varepsilon_\lambda \Delta_\lambda^{-1}$ for the previously described geometries as a function of the number of quantum gates necessary to encode $v_k$ are shown in Fig. 5. A distinction is made between the plain computation as described in the algorithm in the left and an optimized version in the right graph. For the optimization, a compilation and error mitigation method[2] is used, provided by IBM's Qiskit Primitives. This shows the full potential of the device. More details about error mitigation methods are gathered by Cai et al. [25].

---

[2] Here, optimization level=3 and probabilistic error cancellation (resilience level=3) is used. More information at
https://docs.quantum.ibm.com/run/configure-runtime-compilation and
https://docs.quantum.ibm.com/run/configure-error-mitigationnote

*4.2. Results*

The algorithm proposed achieves the ideal resonance frequencies for minimum gate numbers. However, the effect of noise dominates fast and the results are only after ∼ 100 gates closer to the random results of fully mixed states than to the desired values. The gate number increases for all geometries with the system size represented in the required qubit number. This shows that the system size is more important for the accuracy than the form of the geometry. Even state-of-the-art optimization methods[2] do not manage to mitigate the effect of noise drastically. It seems like the simple geometry of the linear coupled oscillator deals slightly better with optimization methods. However, due to the increase in statistical errors from the error mitigation, this is not guaranteed and larger sample sizes are necessary to achieve the same statistical accuracy. Currently, available quantum computers are still affected by noise, making long computations impossible. Nevertheless, it was possible to estimate the resonance frequencies of small industry-related problems. Ongoing improvements in the accuracy of quantum hardware will increase the number of quantum gates that can be executed and therefore the size of the problems that can be addressed. However, the presented hybrid subroutine cannot achieve a quantum advantage over classical algorithms because it computes $g_i, \forall i$ in step 2, which is time-consuming. More advanced algorithms such as QPE, which struggles with input-output bottlenecks as well, are required for this.

## 5. Summary & Outlook

The presented paper introduces a methodical approach to evaluate the potential of QC in the category of manufacturing simulation in the application domain production. The defined QPIs and the developed approach for identifying bottlenecks in manufacturing simulation for which the application of QC potentially provides a speed-up or an increase in accuracy are applied to a milling dynamics simulation. The first results of testing a hybrid routine by using different geometries of coupled oscillators as an application approach for the milling dynamics simulation on a superconducting quantum computer from IBM were presented. However, available quantum devices are still affected by noise which makes long computations impossible. The method presented is not yet complete and additional QPIs may be added to the list, as well as additional steps, as needed. Ongoing improvements, especially in the mentioned hardware-related QPIs will increase the size of the problems that can be addressed. Additionally, efforts will be directed toward scaling up the minimum value problem of coupled oscillators to a three-dimensional approach. To achieve a practical industrial use case, such as the milling dynamics simulation of thin-walled aerospace components, further development of a QPE algorithm will continue in the future.

**Acknowledgements**

This paper has received funding from the 'QUASIM' (01MQ22001A) research, a program of the German Federal Ministry of Economics and Climate Protection (BMWK).